\documentclass[prd,showpacs,twocolumn,floatfix]{revtex4}

\usepackage{graphicx}% Include figure files

\begin{document}

\title{Global analysis of muon decay measurements}

\author{C.A.~Gagliardi}
\affiliation{Cyclotron Institute, Texas A\&M University, College Station, TX 77843}

\author{R.E.~Tribble}
\affiliation{Cyclotron Institute, Texas A\&M University, College Station, TX 77843}

\author{N.J.~Williams}
\altaffiliation[Present address: ]{Langston University, Langston, OK 73050.}
\affiliation{Cyclotron Institute, Texas A\&M University, College Station, TX 77843}

\date{\today}

\begin{abstract}
We have performed a global analysis of muon decay measurements to establish model-independent limits on the space-time structure of the muon decay matrix element.  We find limits on the scalar, vector and tensor coupling of right- and left-handed muons to right- and left-handed electrons.  The limits on those terms that involve the decay of right-handed muons to left-handed electrons are more restrictive than in previous global analyses, while the limits on the other non-standard model interactions are comparable.  The value of the Michel parameter $\eta$ found in the global analysis is $-0.0036 \pm 0.0069$, slightly more precise than the value found in a more restrictive analysis of a recent measurement.  This has implications for the Fermi coupling constant $G_F$.
\end{abstract}

% PACS codes
                                    % Use showpacs class option to display
%\keywords{Suggested keywords}%Use showkeys class option if keyword
                              %display desired
\maketitle

\section{Introduction}

Muon decay, $\mu \to e\nu\bar{\nu}$, is an excellent laboratory to investigate the weak interaction.  The energy and angular distributions of the electrons emitted in polarized muon decay are specified by the muon decay parameters $\rho$, $\eta$, $\xi$, and $\delta$ \cite{Michel,Kino57}, conventionally referred to as the Michel parameters.  The additional decay parameters $\xi'$ and $\xi''$ determine the longitudinal polarization of the outgoing electrons, and the parameters $\alpha$, $\beta$, $\alpha'$, and $\beta'$ determine the transverse polarization \cite{Kino57}.  Radiative muon decay, $\mu \to e\nu\bar{\nu}\gamma$, provides access to another decay parameter, $\bar{\eta}$ \cite{etabar}.  These 11 decay parameters, which are not all independent, together with the muon lifetime provide a complete description of muon decay if the neutrinos are not observed.

The decay parameters are related to the underlying space-time structure of muon decay by the most general, local, derivative-free, Lorentz-invariant transition matrix element, which can be written in terms of helicity-preserving amplitudes as \cite{Fets86}:
\begin{equation}
M = \frac{4 G_F}{\sqrt{2}} \sum_{\gamma=S,V,T;\epsilon,\mu=R,L} g^{\gamma}_{\epsilon\mu}
\langle \bar{e}_{\epsilon} | \Gamma^{\gamma} | \nu_e \rangle
\langle \bar{\nu}_{\mu} | \Gamma_{\gamma} | \mu_{\mu} \rangle .
\label{Eq:mat_elt}
\end{equation}
In this equation, $G_F$ is the Fermi coupling constant, determined from the muon lifetime, and the $g^{\gamma}_{\epsilon\mu}$ specify the scalar, vector and tensor couplings between $\mu$-handed muons and $\epsilon$-handed electrons.  Deviations from this expression due to the non-local nature of the standard model weak interaction are $O(m_{\mu}^2/m_W^2)$ and, thus, are negligible compared to our current experimental knowledge of the muon decay parameters.  Equation (\ref{Eq:mat_elt}) is particularly convenient because the standard model, with pure $V-A$ coupling, implies $g^V_{LL}=1$ and all the other coupling constants are zero.  In contrast, additional coupling constants are non-zero in many extensions to the standard model.  For example, $g^V_{RR}$, $g^V_{LR}$, and $g^V_{RL}$ are non-zero in left-right symmetric models \cite{Herc86}.  For recent reviews of muon decay, see \cite{Fets95,Kuno01,Fets_in_PDG}.

The coupling constants $g^{\gamma}_{\epsilon\mu}$ represent 18 independent real parameters, so it is not possible to determine all of them from the existing muon decay observables.  A global analysis of all the existing muon decay measurements can nonetheless set stringent model-independent experimental limits on the non-standard model contributions \cite{Fets86,Burk85,Stok86,Balk88} that can then be compared to model calculations such as those in \cite{Prez05}, which predict that $g^{S,V,T}_{LR}$ and $g^{S,V,T}_{RL}$ should be very small based on their contributions to neutrino masses.  Recently, new measurements have been performed for the Michel parameters $\rho$ \cite{Muss05} and $\delta$ \cite{Gapo05} and for the transverse polarization parameters $\eta$, $\eta''$, $\alpha'$, and $\beta'$ \cite{Dann05}.  ($\eta$ and $\eta''$ parametrize the momentum-dependence of the $CP$-allowed transverse polarization in a manner different from, but equivalent to, $\alpha$ and $\beta$.  See Eq.\@ (\ref{Eq:params}) below.)  Each of these new results is a factor of $\sim$\,2.5 more precise than the previous best measurement \cite{PDG04}.  This makes a new global analysis timely.  In this paper, we present such an analysis to establish updated limits on possible non-standard model contributions to muon decay.

\section{Fit Procedure}

Several different parametrizations for muon decay have been introduced over the years.  One common version \cite{Kino57}, which is based on the expression for the muon decay matrix element in charge-retention order, describes the muon decay observables in terms of 10 constants: $a/A$, $a'/A$, $b/A$, $b'/A$, $c/A$, $c'/A$, $\alpha/A$, $\beta/A$, $\alpha'/A$, $\beta'/A$.  Each of these 10 constants is a bilinear combination of the coupling constants $g^{\gamma}_{\epsilon\mu}$ \cite{Fets_in_PDG}.  The normalization factor, $A = a+4b+6c = 16$, is fixed by adjusting $G_F$ to reproduce the muon lifetime.  The 9 remaining linearly independent terms may be determined from the experimental values of the muon decay parameters.  This procedure was used in the global analysis performed in \cite{Burk85}.

An alternative parametrization \cite{Fets86} utilizes a different set of bilinear combinations of the coupling constants:
\begin{eqnarray}
Q_{RR} & = & \frac{1}{4}|g^S_{RR}|^2 + |g^V_{RR}|^2 , \nonumber \\
Q_{LR} & = & \frac{1}{4}|g^S_{LR}|^2 + |g^V_{LR}|^2 + 3|g^T_{LR}|^2 , \nonumber \\
Q_{RL} & = & \frac{1}{4}|g^S_{RL}|^2 + |g^V_{RL}|^2 + 3|g^T_{RL}|^2 , \nonumber \\
Q_{LL} & = & \frac{1}{4}|g^S_{LL}|^2 + |g^V_{LL}|^2 , \nonumber \\
B_{LR} & = & \frac{1}{16}|g^S_{LR}+6g^T_{LR}|^2 + |g^V_{LR}|^2 , \nonumber \\
B_{RL} & = & \frac{1}{16}|g^S_{RL}+6g^T_{RL}|^2 + |g^V_{RL}|^2 , \nonumber \\
I_{\alpha} & = & \frac{1}{4}[g^V_{LR}(g^S_{RL}+6g^T_{RL})^* + (g^{V}_{RL})^*(g^S_{LR}+6g^T_{LR})] \nonumber \\
       & = & (\alpha + i\alpha')/2A , \nonumber \\
I_{\beta}  & = & \frac{1}{2}[g^V_{LL}(g^{S}_{RR})^* + (g^{V}_{RR})^*g^S_{LL}] \nonumber \\
       & = & -2(\beta + i\beta')/A .
\label{Eq:bilins}
\end{eqnarray}
These bilinear combinations satisfy the constraints \cite{Fets86}:
\begin{eqnarray}
0 \leq & Q_{\epsilon\mu} & \leq 1, {\rm ~~where~} \epsilon,\mu = R,L , \nonumber \\
0 \leq & B_{\epsilon\mu} & \leq Q_{\epsilon\mu}, {\rm ~~where~} \epsilon\mu = RL, LR , \nonumber \\
|I_{\alpha}|^2 & \leq & B_{LR} B_{RL} , \nonumber \\
|I_{\beta}|^2  & \leq & Q_{LL} Q_{RR} ,
\label{Eq:constraints}
\end{eqnarray}
and the normalization condition,
\begin{equation}
Q_{RR} + Q_{LR} + Q_{RL} + Q_{LL} = 1 .
\label{Eq:norm}
\end{equation}
The advantages of this parametrization are that it contains the maximum number of positive semi-definite bilinear combinations of the $g^{\gamma}_{\epsilon\mu}$ and that the $Q_{\epsilon\mu}$ are directly interpretable as the total probabilities for $\mu$-handed muons to decay into $\epsilon$-handed electrons \cite{Fets86}.  Furthermore, experimentally it is found that $Q_{LL}$ is close to unity.  Thus, if Eq.\@ (\ref{Eq:norm}) is used to eliminate $Q_{LL}$, the 9 remaining variables are all close to zero.  The bilinear combinations $Q_{RR}$, $Q_{LR}$, $Q_{RL}$, $B_{LR}$, $B_{RL}$, $I_{\alpha}$, $I_{\beta}$ were adopted in the global analyses presented in \cite{Fets86,Stok86,Balk88}.

For convenience, in the present analysis we have adopted a hybrid set of independent variables:  $Q_{RR}$, $Q_{LR}$, $Q_{RL}$, $B_{LR}$, $B_{RL}$, $\alpha/A$, $\beta/A$, $\alpha'/A$, $\beta'/A$.  The muon decay parameters may be expressed in terms of these variables as \cite{Fets95,Fets_in_PDG}:
\begin{eqnarray}
\rho & = & \frac{3}{4} + \frac{1}{4}(Q_{LR}+Q_{RL}) - (B_{LR}+B_{RL}) , \nonumber \\
\xi  & = & 1 - 2Q_{RR} - \frac{10}{3}Q_{LR} + \frac{4}{3}Q_{RL} + \frac{16}{3}(B_{LR}-B_{RL}) , \nonumber \\
\xi \delta & = & \frac{3}{4} - \frac{3}{2}Q_{RR} - \frac{7}{4}Q_{LR} + \frac{1}{4}Q_{RL} + (B_{LR}-B_{RL}) , \nonumber \\
\xi' & = & 1 - 2Q_{RR} - 2Q_{RL} , \nonumber \\
\xi'' & = & 1 - \frac{10}{3}(Q_{LR}+Q_{RL}) + \frac{16}{3}(B_{LR}+B_{RL}) , \nonumber \\
\bar{\eta} & = & \frac{1}{3}(Q_{LR}+Q_{RL}) + \frac{2}{3}(B_{LR}+B_{RL}) , \nonumber \\
\eta & = & (\alpha - 2\beta)/A , \nonumber \\
\eta'' & = & (3\alpha + 2\beta)/A .
\label{Eq:params}
\end{eqnarray}

We have computed the joint probability density function of the 9 independent variables using Monte Carlo integration techniques, in a manner similar to that described in \cite{Burk85}.  For each experimental input, we have assumed that the corresponding probability distribution takes the form of a one- or two-dimensional Gaussian, truncated to the allowed parameter region if necessary.  When statistical and systematic uncertainties have been quoted separately, we have added them in quadrature.

\section{Input Values}

Each of the previous muon decay global analyses \cite{Fets86,Burk85,Stok86,Balk88} utilized the same input parameters -- $\rho$, $\delta$, $P_{\mu}\xi\delta/\rho$, $\xi'$, $\xi''$, $\alpha/A$, $\beta/A$, $\alpha'/A$, $\beta'/A$.  For each of these quantities, the previous analyses adopted the single most precise experimental measurement that was available at the time.

\begin{table}[tb]
\caption{\label{Tab:inputs}
Experimental measurements included in the global analysis.}
\begin{ruledtabular}
\begin{tabular}{ccc}
Parameter     &      Value      & Reference \\
\hline
$\rho$        & 0.7518 $\pm$ 0.0026   & \cite{PDG04}    \\
              & 0.75080 $\pm$ 0.00105\footnotemark[1] & \cite{Muss05}   \\
$\delta$      & 0.7486 $\pm$ 0.0038   & \cite{Balk88}   \\
              & 0.74964 $\pm$ 0.00130 & \cite{Gapo05}   \\
$P_{\mu}\xi$  & 1.0027 $\pm$ 0.0085\footnotemark[2]   & \cite{Belt87}   \\
$P_{\mu}\xi\delta/\rho$  & 0.99787 $\pm$ 0.00082\footnotemark[2] & \cite{Jodi86,Jodi88}  \\
$\xi'$        & 1.00 $\pm$ 0.04       & \cite{PDG04}    \\
$\xi''$       & 0.65 $\pm$ 0.36       & \cite{Burk85a}  \\
$\bar{\eta}$  & 0.02 $\pm$ 0.08       & \cite{PDG04}    \\
$\alpha/A$    & 0.015 $\pm$ 0.052\footnotemark[3]     & \cite{Burk85}   \\
$\beta/A$     & 0.002 $\pm$ 0.018\footnotemark[3]     & \cite{Burk85}   \\
$\eta$        & 0.071 $\pm$ 0.037\footnotemark[4]     & \cite{Dann05}   \\
$\eta''$      & 0.105 $\pm$ 0.052\footnotemark[4]     & \cite{Dann05}   \\
$\alpha'/A$   & -0.047 $\pm$ 0.052\footnotemark[5]    & \cite{Burk85}   \\
              & -0.0034 $\pm$ 0.0219\footnotemark[6]  & \cite{Dann05}   \\
$\beta'/A$    & 0.017 $\pm$ 0.018\footnotemark[5]     & \cite{Burk85}   \\
              & -0.0005 $\pm$ 0.0080\footnotemark[6]  & \cite{Dann05}   \\
\end{tabular}
\end{ruledtabular}
\footnotetext[1]{\,Correlated with $\eta$ in \protect\cite{Muss05}.  See text.}
\footnotetext[2]{\,Fit assumes $P_{\mu} = 1$.  See text.}
\footnotetext[3]{\,$\rho(\alpha/A,\beta/A) = -0.894$ in \protect\cite{Burk85}.}
\footnotetext[4]{\,$\rho(\eta,\eta'') = +0.946$ in \protect\cite{Dann05}.}
\footnotetext[5]{\,$\rho(\alpha'/A,\beta'/A) = -0.894$ in \protect\cite{Burk85}.}
\footnotetext[6]{\,$\rho(\alpha'/A,\beta'/A) = -0.893$ in \protect\cite{Dann05}.}
\end{table}

We have utilized a different philosophy.  We include all of the accepted muon decay parameter measurements that are reported in \cite{PDG04}, with two exceptions.  When \cite{PDG04} determines a decay constant from a single input, we have included it.  When \cite{PDG04} computes an average from a set of previous measurements, we have utilized this average value in our fits.  The exceptions involve the values adopted in \cite{PDG04} for $\eta$ and the transverse polarization parameters $\alpha/A$, $\beta/A$, $\alpha'/A$, and $\beta'/A$.  The adopted values for these five parameters are all derived from the global fit in \cite{Burk85}.  Rather, for the transverse polarization parameters, we have used the experimental results \cite{Burk85} that were inputs to that previous global analysis.  Meanwhile, we have not included the previous measurement of $\eta$ in \cite{Deren69}.  Its uncertainty is a factor of 30 larger than the uncertainty in $\eta$ derived from our fit to the rest of the existing data.

In addition to the measurements reported in \cite{PDG04}, we have included the recent measurements of the Michel parameters $\rho$ \cite{Muss05} and $\delta$ \cite{Gapo05} and of the transverse polarization parameters $\eta$, $\eta''$, $\alpha'/A$, and $\beta'/A$ \cite{Dann05}.  The complete list of inputs is given in Table \ref{Tab:inputs}.

Three of the input values require special consideration.  In \cite{Muss05}, the value obtained for $\rho$ assumed that $\eta$ is given by the accepted value, $-0.007$, and a contribution of $\pm$\,0.00023 was assigned to the uncertainty in $\rho$ associated with the $\pm$\,0.013 uncertainty in the accepted value for $\eta$ \cite{PDG04}.  We have performed two separate global fits.  In one fit, we included the $\pm$\,0.00023 in the uncertainty for $\rho$, then treated the measurement as uncorrelated with $\eta$.  In the other fit, we included the dependence of the measured value of $\rho$ on $\eta$ \cite{Muss05} and reduced the uncertainty in $\rho$ accordingly.  The two fits give very similar results.  The results for the latter are reported here.

References \cite{Belt87} and \cite{Jodi86,Jodi88} report measurements of $P_{\mu}\xi$ and $P_{\mu}\xi\delta/\rho$, respectively.  In each case, $P_{\mu}$ represents the polarization of the muon in pion decay.  This is known to be $>$\,0.99682 with 90\% confidence \cite{Jodi86,Jodi88,Fets84}.  We have assumed $P_{\mu} = 1$ when including the $P_{\mu}\xi$ measurement in our fits.  In contrast, \cite{Jodi86,Jodi88} specifically mention the existence of possible unknown sources of muon depolarization prior to the decay and, therefore, choose to interpret the measurement of $P_{\mu}\xi\delta/\rho$ as a lower limit.  In recognition of this, we have performed two separate fits.  In one fit, we assumed the probability distribution for $\xi\delta/\rho$ is Gaussian as given in Table \ref{Tab:inputs}, which is equivalent to assuming $P_{\mu} = 1$.  In the other, we treated the probability distribution as Gaussian when $\xi\delta/\rho$ is less than the central value and constant when $\xi\delta/\rho$ is greater than the central value.  We report results from the former fit because it finds slightly less restrictive limits for $Q_{RR}$, $Q_{LR}$, and $B_{LR}$, while the remaining parameters are unchanged between the two fits.

\section{Results}

Table \ref{Tab:results} shows the results of the global fits.  90\% confidence level upper limits (lower limit for $Q_{LL}$) are given for the positive semi-definite quantities.  For completeness, mean and r.m.s.\@ values are also specified for these variables, even though the corresponding probability distributions are far from Gaussian.  In contrast, the output probability distributions for $\alpha/A$, $\beta/A$, $\alpha'/A$, and $\beta'/A$ are very close to Gaussian.

\begin{table}[tb]
\caption{\label{Tab:results}
Results of the global fit.  For the positive semi-definite quantitites, 90\% confidence level upper limits (lower limit for $Q_{LL}$) are given, and for completeness, the mean and r.m.s.\@ values from the fit are also given in parentheses.}
\begin{ruledtabular}
\begin{tabular}{cc}
Parameter   & Fit Result ($\times 10^3$) \\
\hline
$Q_{RR}$    & $<$\,1.14 ~(0.60 $\pm$ 0.38) \\
$Q_{LR}$    & $<$\,1.94 ~(1.22 $\pm$ 0.53) \\
$B_{LR}$    & $<$\,1.27 ~(0.72 $\pm$ 0.40) \\
$Q_{RL}$    & $<$\,44 ~(26 $\pm$ 13)       \\
$B_{RL}$    & $<$\,10.9 ~(6.4 $\pm$ 3.3)   \\
$Q_{LL}$    & $>$\,955 ~(973 $\pm$ 13)     \\
$\alpha/A$  & 0.3 $\pm$ 2.1   \\
$\beta/A$   & 2.0 $\pm$ 3.1   \\
$\alpha'/A$ & -0.1 $\pm$ 2.2  \\
$\beta'/A$  & -0.8 $\pm$ 3.2  \\
\end{tabular}
\end{ruledtabular}
\end{table}

Only a subset of the input quantities play a role in constraining each of the fit parameters.  $Q_{RR}$ is constrained primarily by the measurement of $P_{\mu}\xi\delta/\rho$.  $Q_{LR}$ and $B_{LR}$ are constrained by the measurements of $\rho$, $\delta$, and $P_{\mu}\xi\delta/\rho$.  $Q_{RL}$ is constrained by the measurement of $\xi'$.  $B_{RL}$ is constrained by the combination of $\xi'$, $\rho$, and $\delta$.  $\alpha/A$ and $\alpha'/A$ are constrained primarily by the requirement $|I_{\alpha}|^2 \leq B_{LR}B_{RL}$.  Finally, the large reduction in the uncertainties for $\alpha/A$ and $\alpha'/A$ obtained when applying the $I_{\alpha}$ constraint significantly reduces the correlations between $\alpha/A$ and $\beta/A$ and between $\alpha'/A$ and $\beta'/A$.  The fit finds $\rho(\alpha/A,\beta/A) = -0.19$ and $\rho(\alpha'/A,\beta'/A) = -0.20$.  The correlations dominate the uncertainties in $\beta/A$ and $\beta'/A$ quoted in Table \ref{Tab:inputs}, so the result is that the transverse polarization measurements become far more precise.  Overall, the constraints in Eq.\@ (\ref{Eq:constraints}) play a crucial role in reducing the uncertainties for a number of the fitted parameters, as was noted in \cite{Burk85}.

The fit results may be used to determine limits on the coupling constants $g^{\gamma}_{\epsilon\mu}$.  These are given in Table \ref{Tab:limits}.  Limits are also given for certain linear combinations of scalar and tensor interactions.  Muon decay measurements alone are unable to separate contributions of $g^S_{LL}$ from the standard model term, $g^V_{LL}$.  Other experimental results, such as the branching ratio for $\pi \to e\nu$, argue that the scalar contribution must be very small, but this requires assumptions that go beyond the model-independent experimental limits that we are exploring here.  In \cite{Fets86} it was noted that the rate for inverse muon decay, $\nu_{\mu} e^- \to \mu^- \nu_e$, may be used to distinguish between $g^S_{LL}$ and $g^V_{LL}$.  In Table \ref{Tab:limits}, the limits quoted for $g^S_{LL}$ and $g^V_{LL}$ come from \cite{Fets_in_PDG}, which utilized the inverse muon decay measurements in \cite{Vila95}.

\begin{table}[tb]
\caption{\label{Tab:limits}
90\% confidence limits on the muon decay coupling constants in Eq.\@ (\ref{Eq:mat_elt}) are compared to the previous accepted values from \protect\cite{Fets_in_PDG}.  Limits are also given for certain scalar-tensor interference combinations.}
\begin{ruledtabular}
\begin{tabular}{ccc}
  &  Ref.\@ \protect\cite{Fets_in_PDG} & Present work \\
\hline
$|g^S_{RR}|$  & $<$\,0.066  & $<$\,0.067   \\
$|g^V_{RR}|$  & $<$\,0.033  & $<$\,0.034   \\
$|g^S_{LR}|$  & $<$\,0.125  & $<$\,0.088   \\
$|g^V_{LR}|$  & $<$\,0.060  & $<$\,0.036   \\
$|g^T_{LR}|$  & $<$\,0.036  & $<$\,0.025   \\
$|g^S_{RL}|$  & $<$\,0.424  & $<$\,0.417   \\
$|g^V_{RL}|$  & $<$\,0.110  & $<$\,0.104   \\
$|g^T_{RL}|$  & $<$\,0.122  & $<$\,0.104   \\
$|g^S_{LL}|$  & $<$\,0.550  & $<$\,0.550   \\
$|g^V_{LL}|$  & $>$\,0.960  & $>$\,0.960   \\
\hline
$|g^S_{LR}+6g^T_{LR}|$  &   & $<$\,0.143   \\
$|g^S_{LR}+2g^T_{LR}|$  &   & $<$\,0.108   \\
$|g^S_{LR}-2g^T_{LR}|$  &   & $<$\,0.070   \\
$|g^S_{RL}+6g^T_{RL}|$  &   & $<$\,0.418   \\
$|g^S_{RL}+2g^T_{RL}|$  &   & $<$\,0.417   \\
$|g^S_{RL}-2g^T_{RL}|$  &   & $<$\,0.418   \\
\end{tabular}
\end{ruledtabular}
\end{table}

Table \ref{Tab:limits} shows that the present limits on $|g^S_{LR}|$, $|g^V_{LR}|$, and $|g^T_{LR}|$ are significantly better than those in \cite{Fets_in_PDG}, which arises from the inclusion of the new measurements of $\rho$ and $\delta$.  This was already noted in \cite{Gapo05}, where calculations based on the measurements of $\rho$, $\delta$, and $P_{\mu}\xi\delta/\rho$ gave a slightly more restrictive limit for $|g^S_{LR}|$ and a less restrictive limit for $|g^V_{LR}|$.  The new limits for $|g^S_{RR}|$ and $|g^V_{RR}|$ in Table \ref{Tab:limits} are slightly \textit{less restrictive} than their previously accepted values \cite{Fets_in_PDG}.  In \cite{Fets_in_PDG}, these limits came from calculations in \cite{Balk88} that were performed before an error in the reported value of $P_{\mu}\xi\delta/\rho$ \cite{Jodi86} had been discovered.  The present fit utilizes the corrected, less restrictive value of $P_{\mu}\xi\delta/\rho$ \cite{Jodi88}.  The new results for $|g^S_{RL}|$ and $|g^V_{RL}|$ are little different from the previously accepted values since the primary experimental input that constrains these, $\xi'$, has not been remeasured since \cite{Burk85a}.

The Michel parameter $\eta$ merits special discussion because it has an impact on $G_F$.  $G_F$ is typically calculated from the measured muon lifetime assuming a pure $V-A$ interaction.  This leads to the quoted precision of $\Delta G_F/G_F = 9 \times 10^{-6}$ \cite{PDG04}.  However, a non-zero value of $\eta$ would change the muon decay phase space so that \cite{Sche78}:
\begin{equation}
G_F \approx G_F^{(V-A)} (1 - 2\eta m_e/m_{\mu}) .
\label{Eq:G_F}
\end{equation}
Thus, the $\pm$\,0.013 uncertainty in the accepted value of $\eta$ \cite{PDG04} leads to an additional uncertainty in $G_F$ of $\Delta G_F/G_F = 1.3 \times 10^{-4}$.  The recent measurement of the transverse polarization of the electrons emitted in muon decay \cite{Dann05} included two analyses of the results.  The general (model-independent) analysis led to the values of $\eta$, $\eta''$, $\alpha'/A$, and $\beta'/A$ quoted in Table \ref{Tab:inputs}.  A second, restricted analysis assumed that muon decay can be described with only two coupling constants, $g^V_{LL}$ and $g^S_{RR}$.  Equations (\ref{Eq:bilins}) and (\ref{Eq:params}) imply that this assumption requires $\alpha/A = \alpha'/A = 0$ and $\eta'' = -\eta$.  The restricted analysis concluded $\eta = -0.0021 \pm 0.0070 \pm 0.0010$ and $\beta'/A = -0.0013 \pm 0.0035 \pm 0.0006$.  As noted above, the allowed ranges of $\alpha/A$ and $\alpha'/A$ are severely constrained by other muon decay data in the present global analysis.  Thus, we find a model-independent result, $\eta = -0.0036 \pm 0.0069$, with slightly better precision than that of the model-dependent restricted analysis in \cite{Dann05}.  This reduces the contribution of $\eta$ to the uncertainty in $G_F$ to $\Delta G_F/G_F = 6.7 \times 10^{-5}$.  Note that all of the new measurements \cite{Muss05,Gapo05,Dann05} play important roles in reducing the uncertainty in $\eta$.

\section{Conclusion}

We have performed a new global analysis of all the existing data on muon decay that do not include observation of the outgoing neutrinos, including recent measurements of the Michel parameters $\rho$ and $\delta$ and the muon decay transverse polarization parameters $\eta$, $\eta''$, $\alpha'/A$, and $\beta'/A$.  The global analysis finds that the upper limits on the coupling constants $|g^S_{LR}|$, $|g^V_{LR}|$, and $|g^T_{LR}|$ are more restrictive than the previous accepted values.  It also finds that $\eta = -0.0036 \pm 0.0069$, which is nearly a factor of two more precise than the previous accepted value.

\acknowledgments

We thank colleagues in the TWIST Collaboration for many useful discussions regarding muon decay limits.  This work was supported in part by the U.S. National Science Foundation Research Experiences for Undergraduates program, grant number 0354098, and by the U.S. Department of Energy, grant number DE-FG03-93ER40765.

\end{document}